\documentclass[journal=jacsat,manuscript=article]{achemso}

\usepackage{chemformula} 
\usepackage[T1]{fontenc} 
\usepackage{amssymb}

\usepackage{cases}

\newcommand{\md}{\mathrm{d}}

\newcommand{\G}{\mathbb{G}}

\author{Xiaomei Yao}
\affiliation{Beijing International Center for Mathematical Research, Peking University, Beijing
100871, China.}

\author{Jie Xu}
\affiliation{LSEC \& NCMIS, Institute of Computational Mathematics and Scientific/Engineering Computing (ICMSEC), Academy of Mathematics and Systems Science (AMSS), Chinese Academy of Sciences, Beijing, China}
\email{xujie@lsec.cc.ac.cn}

\author{Lei Zhang}
\affiliation{Beijing International Center for Mathematical Research, Peking University, Beijing
100871, China.}
\email{zhangl@math.pku.edu.cn}

\title[interface transition]
  {Transition pathways in Cylinder-Gyroid interface}


\begin{document}


\begin{abstract}
When two distinct ordered phases contact, the interface may exhibit rich and fascinating structures. Focusing on the Cylinder-Gyroid interface system, transition pathways connecting various interface morphologies are studied armed with the Landau--Brazovskii model. Specifically, minimum energy paths are obtained by computing transition states with the saddle dynamics. We present four primary transition pathways connecting different local minima, representing four different mechanisms of the formation of the Cylinder-Gyroid interface. The connection of Cylinder and Gyroid can be either direct or indirect via Fddd with three different orientations. Under different displacements, each of the four pathways may have the lowest energy.
\end{abstract}

\section{Introduction}

Modulated phases of similar patterns can be formed by block copolymer melts\cite{leibler1980theory,Thomas1988} and many totally distinct materials, such as biological cells\cite{Luzzati1993Cubic} and metal nanoparticles\cite{Warren2008Ordered,Song2020Oriented}.
Of these modulated phases, the most commonly observed patterns include Lamellae(L), Cylinder(C), Sphere(BCC, FCC), Gyroid (G) and Fddd, for which
extensive studies have been carried out both experimentally\cite{Orilall2011Block,Mai2012Self,Stefik2015Block} and theoretically\cite{Matsen1994Stable,Tyler2005Orthorhombic,Jiang2013Discovery}.
These phases, while possessing distinct symmetries, can coexist in many cases. The interface between two phases would exhibit fascinating structures, which also characterize 
first-order phase transitions\cite{Kumar2018Why,Cao2016Interconversion,Bao2021Discovery}.

Because of the intrinsic ordered structures, their relative positions and orientations are essential to the interface, which is evidenced by several epitaxial relations\cite{Bang2003Mechanisms,Park2009New,Vukovic2012Double,Jung2014Epitaxial}.
In particular, multiple interface morphologies and epitaxies are found for the C-G coexistence systems\cite{Park2009New}, which result in distinct processes of phase transitions. These experimental findings bring us a number of glamorous phenomena, meanwhile raise theoretical problems on the formation mechanisms that require enlightening perspectives.

Some theoretical attempts have been made to understand the underlying mechanism of interfaces.
One convenient approach is to follow relaxation dynamics, typically carried out in a large cell, to let interface emerge and evolve\cite{Rogers1989,Wickham2003Nucleation,Elder2004Modeling}.
Usually, the results from the dynamic approach contain multiple interfaces that might interact one another.
Furthermore, it is not easy to fix the relative orientation and displacement in large cell simulations.
To arrive at closer examinations of the interfacial structures, some posed two phases delicately in a small computational box with special orientations and displacements.
Using this approach, grain boundaries of L\cite{Netz1997Interfaces,Tsori2000Defects,Duque2002Theory},
BCC \cite{Jaatinen2009Thermodynamics} and cubic phases\cite{Belushkin2009Twist} were examined.

Interfaces of general relative positions and orientations are also studied. For example, to provide insights on epitaxy, an artificial mixing ansatz is adopted followed by searching the minimum exceeding energy\cite{McMullen1988The,Yamada2007Interface,Wang2011Origin}. 
However, the interfacial morphology obtained from this approach may be far from optimal in many cases.
A framework dealing with general relative positions and orientations was proposed later \cite{Xu2017Computing}, where the boundary conditions and basis functions are carefully chosen to fix the bulk phases at certain positions and orientations consistently.
This framework is further equipped with delicate numerical methods that can successfully deal with quasiperiodic interface \cite{Cao2020Computing}.
The interfacial structures obtained from this framework prove to be much more complicated than simple mixing.
Even in the simplest cases, a series of energy minima can occur depicting the process of phase transition.
Moreover, when we alter the relative positions and orientations, a few fascinating results are then obtained, implying that the underlying mechanism can be quite complex in the formation of interfaces, such as deformation, wetting by a third phase, zigzagging, etc.

The above results indicate that the interface system could possess multiple energy minima.
The relationships between minima can be characterized by the minimum energy paths (MEPs) on the free-energy landscape, which represent the most probable transition pathways \cite{E2002String}.
The crest of a MEP connecting two minima is regarded as the transition state that is an (Morse) index-1 saddle point.
Thus, if multiple minima exist, one could imagine that the interface shall be moving along the transition pathways through a series of transition states and minima. 
Nevertheless, the existing results are far from well-understood on the transition pathways in the interface systems.

In this work, we examine the transition pathways connecting different interface morphologies using the Landau--Brazovskii (LB) theory.
Specifically, we apply an efficient numerical method based on the index-1 saddle dynamics to the LB model in order to obtain MEPs connecting various local minima. 
Our focus is the C-G interface system. 
We present four primary transition pathways, representing four different mechanisms of the formation of the C-G interface. The connection of C and G can be either direct or indirect via Fddd with three different orientations. We demonstrate that, when altering the relative positions of C and G, each of the four pathways may have the lowest energy.

The paper is organized as follows. In section 2, we briefly describe the LB model, interface system and the numerical methods for computing the transition pathways. The results are presented in section 3, where we examine transition pathways in the C-G interface systems with different displacements. Discussion and conclusion are given in section 4.

\bigskip
\section{Model and numerical methods}\label{W1}

\subsection{LB free energy and interface system}
The LB model provides a framework for systems that are undergoing microphase separations \cite{Brazovskii1975Phase,Fredrickson1987Fluctuation,Kats1993Weak}, described by the modulation of a scalar $\phi$. 
It has been studied numerically for various modulated phases\cite{Zhang2008An,Mkhonta2010Novel,Wang2011Origin,Spencer2013Simulation,Xu2017Computing}.
The LB free energy density (energy per volume) consists of a term featuring a preferred wavelength and a bulk term given by a quartic polynomial,
\begin{equation}\label{LB}
E[\phi]=\frac{1}{V}\int_\Omega {\rm d}{\bf r}\left\{\frac{\xi^2}{2}[(\nabla^2+q_0^2)\phi]^2+\frac{\tau}{2}\phi^2-\frac{\gamma}{3!}\phi^3+\frac{1}{4!}\phi^4\right\},
\end{equation}
where $q_0=1$ is a characteristic wavelength scale, and $\xi$, $\tau$, $\gamma$ are phenomenological coefficients.
The free energy is combined with the conservation of $\phi$, given by $(1/V) \int_{\Omega}\phi \md {\bf r}=0$.

In general, for a periodic phase $\alpha$, its profile can be obtained by minimizing \eqref{LB} in a unit cell, which can typically be chosen as a cube.
The size of the unit cell can either be estimated \cite{Jiang2013Discovery} or optimized during minimization \cite{Zhang2008An}.
Since we need to utilize the profile of the periodic phases in the C-G interface system, the estimated unit cell is chosen so that two phases can be well matched. Such a choice is also supported by experimental results \cite{Hajduk1994Observation,Bang2003Mechanisms,Park2009New}.

\begin{figure}[!t]
\centering
  \includegraphics[width=0.6\columnwidth]{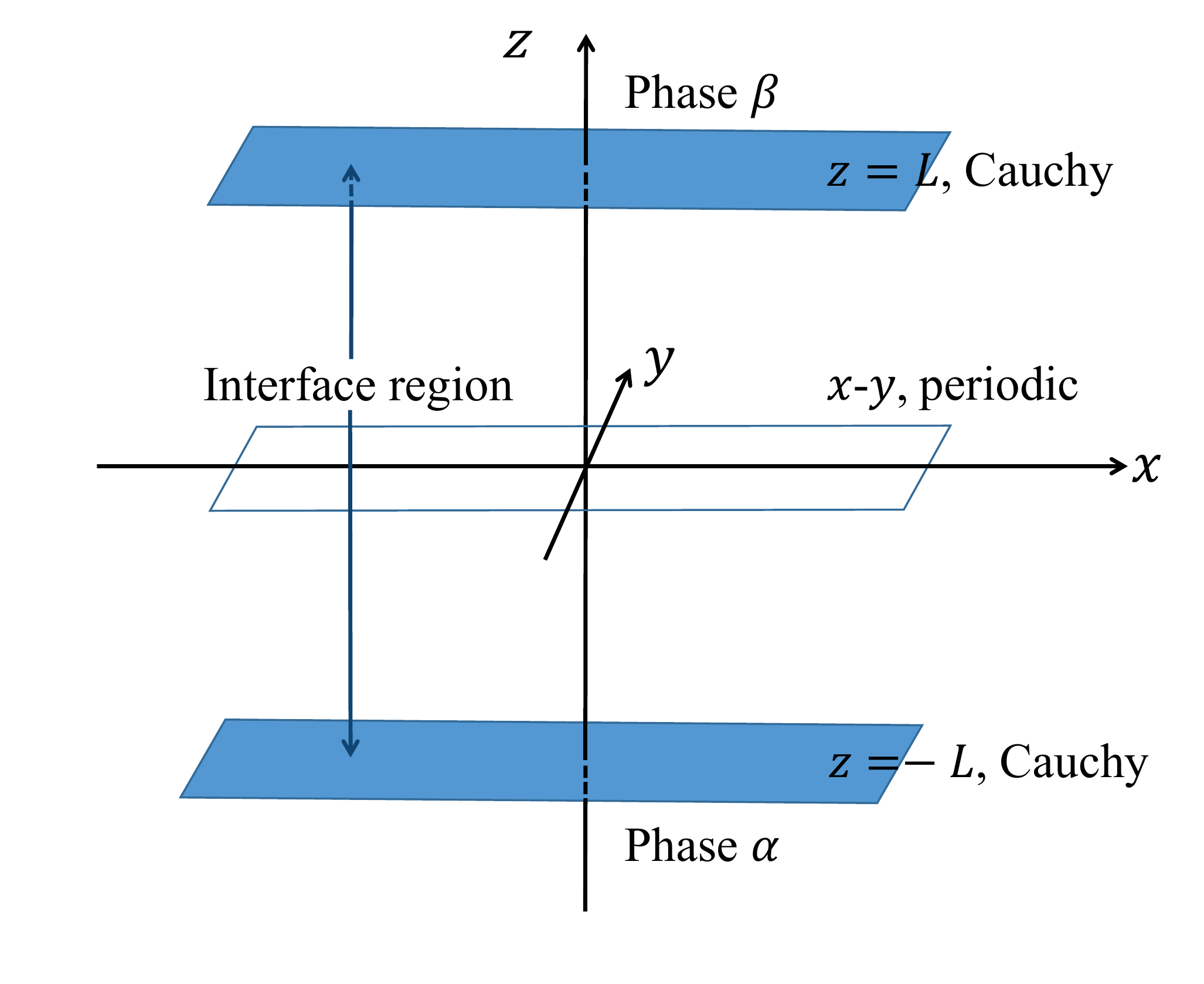}
  \caption{Setting of the interface system. Phase $\alpha$ and phase $\beta$ are lying along $z$-axis, and the system always keep periodic in the $x$-$y$ plane.
    \label{domain}}
\end{figure}

Once the profile of the two phases $\alpha$ and $\beta$ (denoted by $\phi_{\alpha}$ and $\phi_{\beta}$, respectively) are obtained, we can formulate the system for an interface between them, as illustrated in Figure \ref{domain}.
The whole space is divided into three regions by two parallel planes $z =-L$ and $z=L$ for some $L$, with the phase $\alpha$ occupying the region $z\leq -L$ and the phase $\beta$ occupying $z\geq L$. Hence, an interface will form in the interface region $-L<z<L$.
We may manually set $\phi_{\alpha}$ and $\phi_{\beta}$ to be displaced and rotated into the positions and orientations that we desire to pose.
Meanwhile, we assume that, under such positions and orientations, two phases have the same periods $L_x\times L_y$ in the $x$-$y$ plane.
Correspondingly, the computational box in the $x$- and $y$-directions is chosen as $[0,L_x]\times [0,L_y]$ and the periodic boundary conditions are imposed.
Next, a suitable interval $[-L,L]$ is chosen to contain the interfacial region in the $z$-direction, and the information for $z\leq -L$ and $z\geq L$ can be translated into the Cauchy boundary conditions, which are given by the following,
\begin{align*}
&\phi(-L)=\phi_\alpha(-L),\quad\partial_z\phi(-L)=\partial_z\phi_\alpha(-L);\\
&\phi(L)=\phi_\beta(L),\quad\partial_z\phi(L)=\partial_z\phi_\beta(L).
\end{align*}
The conservation of $\phi$ in the computational domain $ [0,L_x]\times [0,L_x]\times [-L,L]$ then becomes
\begin{equation}
\int_0^{L_x} {\rm d}x\int_0^{L_y}{\rm d}y \int_{-L}^L {\rm d}z\, \phi({\bf r})=\int_0^{L_x} {\rm d}x\int_0^{L_y} {\rm d}y\left(\int_{-L}^0 {\rm d}z\, \phi_{\alpha}({\bf r})+\int_0^L {\rm d}z\, \phi_{\beta}({\bf r})\right).
\end{equation}

\subsection{Spatial discretization}
Since the interface system is periodic in the $x$-$y$ plane, we discretize these two directions by Fourier modes. That is, we write $\phi$ as
\begin{equation}
\phi(x,y,z)=\sum_{{\bf G}}\phi_{\bf G}(z)\exp\big(\sqrt{-1}{\bf G}\cdot{\bf r}'\big),
\end{equation}
where ${\bf r}'=(x,y)$, and the reciprocal vectors ${\bf G}=(2\pi m/L_x,2\pi n/L_y)$ with the summation truncated at $|m|, |n|\leq N_p$.
Then, we apply the finite difference method in the $z$-direction.
We define the grid points $z_j=-L+(j-1)\Delta z$, where $\Delta z=2L/(N-1)$, $j=1,\ldots N$.
The second-order derivative about $z$ is discretized as
\begin{equation}
  \partial_z^2\phi(z_j)\approx \frac{z_{j+1}-2z_j+z_{j-1}}{\Delta z^2}.
\end{equation}
The discrete boundary conditions are given by
\begin{align}\label{xbc}
  \phi(z_{0})&=\phi_{\alpha}(z_{0}),\quad\phi(z_1)=\phi_{\alpha}(z_1),\nonumber\\
  \phi(z_N)&=\phi_{\beta}(z_N),\quad \phi(z_{N+1})=\phi_{\beta}(z_{N+1}).
\end{align}

In this way, the free energy density \eqref{LB} can be discretized as a function of the Fourier coefficients $\phi_{\bf G}(z_j)$ at several grid points $z_j$. Such a discretization is also suitable for calculating the phase profiles $\phi_{\alpha}$ and $\phi_{\beta}$ (just substitute the boundary conditions \eqref{xbc} with periodic ones). Thus, we could obtain the phase profiles under the above discretization and set the boundary conditions \eqref{xbc} directly.

The gradient vector is then calculated as
\begin{align}\label{gradient}
  F(\phi_{\bf G}(z_j))=&[\xi^2(-{\bf G}^2+\delta_z^2+1)^2+\tau]\phi_{\bf G}(z_j)-{\gamma\over 2}\sum_{{\bf G}_1+{\bf G}_2={\bf G}}\phi_{{\bf G}_1}(z_j)\phi_{{\bf G}_2}(z_j)\nonumber\\
  &+\frac{1}{6}\sum_{{\bf G}_1+{\bf G}_2+{\bf G}_3={\bf G}}\phi_{{\bf G}_1}(z_j)\phi_{{\bf G}_2}(z_j)\phi_{{\bf G}_3}(z_j)-\lambda\delta({\bf G}=0),
\end{align}
where a projection, with the Lagrange multiplier $\lambda$, on the gradient vector is incorporated to guarantee the conservation of $\phi$.

\subsection{Numerical methods for transition pathways}
It has been noticed that multiple local minima exist in the C-G interface system illustrated above.
Thus, it is natural to investigate the transition pathways connecting different minima via the transition states, which depict how the C-G interface moves. 
There are two classes of numerical methods for the calculation of the transition pathways connecting local minima: chain-of-state methods \cite{E2002String,E2007Simplified,Du2009A} and surface walking methods \cite{Henkelman2000Improved,E2011The,Zhang2012Shrinking,Gao2015An}.
The chain-of-state methods would require knowledge of both initial final states, and a decent initial path connecting them.
The surface walking methods, on the contrary, are able to explore the transition states starting from an initial state.
In the interface system, since we aim to search possible local minima and transition pathways without a priori assumption on the final state, we adopt the index-1 saddle dynamics (SD) to compute the transition states \cite{Zhang2016OSD, yin2019high}, governed by
\begin{subnumcases}{\label{OSD}}
\frac{\partial\phi_{\bf G}}{\partial t}=-(I-2vv^\top){F}(\phi_{\bf G}),\vspace{1mm}\label{OSDa}\\
\frac{\partial v}{\partial t}=-(I-vv^\top)\G (\phi_{\bf G})v,
\end{subnumcases}
where $I$ is the identity matrix, ${F}(\phi_{\bf G})$ is the gradient vector in \eqref{gradient} and $\G$ is the Hessian matrix at $\phi_{\bf G}$. The vector $v$ represents the ascending direction corresponding to the eigenvector of the smallest eigenvalue of the Hessian $\G (\phi_{\bf G})$.
Such SD is built on the fact that an index-1 saddle point is a maximum along the lowest curvature mode and a minimum along all other modes.
In particular, when $v=0$, the SD \eqref{OSDa} leads to the gradient dynamics, which is utilized to search the connected new local minima. 

To avoid direct computation of the Hessian, the shrinking dimer technique is applied to approximate the action of Hessian at the dimer center along the direction $v$, i.e.,
\begin{equation}
\G (\phi_{\bf G})v\approx\frac{{F}(\phi_{\bf G}+lv)-{F}(\phi_{\bf G}-lv)}{2l}\triangleq \frac{F_+-F_-}{2l},
\end{equation}
where the dimer length satisfies ${\md l}/{\md t}=-l$.
The above dynamics is further discretized in time, using a semi-implicit scheme for $\phi_{\bf G}$ and an explicit scheme for $v$,
\begin{subnumcases}{\label{OSD2}}
\phi^{n+1}_{\bf G}=\phi^n_{\bf G}+\Delta t_n\left(F_1(\phi^{n+1}_{\bf G})+F_2(\phi^n_{\bf G})+2v^n(v^n)^\top {F}(\phi^n_{\bf G})\right),\vspace{2mm}\label{OSD2a}\\
v^{n+1}=v^n-\gamma_n\left(\frac{(F_+^n-F_-^n)}{2l_n}-v^n(v^n)^\top\frac{(F_+^n-F_-^n)}{2l_n}\right),\vspace{1mm}\\
l_{n+1}=\frac{l_n}{1+\delta},
\end{subnumcases}
where
\begin{align}
F_1\left(\phi^{n+1}_{\bf G}(z_j)\right)&=-[\xi^2(-{\bf G}^2+\delta_z^2+1)^2+\tau]\phi_{\bf G}^{n+1}(z_j),\nonumber\\
F_2\left(\phi^n_{\bf G}(z_j)\right)&={\gamma\over2}\sum_{{\bf G}_1+{\bf G}_2={\bf G}}\phi_{{\bf G}_1}^n(z_j)\phi_{{\bf G}_2}^n(z_j)-{1\over6}\sum_{{\bf G}_1+{\bf G}_2+{\bf G}_3={\bf G}}\phi_{{\bf G}_1}^n(z_j)\phi_{{\bf G}_2}^n(z_j)\phi_{{\bf G}_3}^n(z_j)\nonumber\\
&+\lambda\delta({\bf G}=0);
\end{align}
$\delta$ is a positive constant to ensure that $l_n\rightarrow 0$ when $n\rightarrow+\infty$, for which we take $\delta=0.01$ in the simulation; the time steps $\Delta t_n$ is chosen adaptively by \cite{Qiao2011An,Cao2020Computing}
\begin{equation}\label{adaptive}
\Delta t_{n}=\max\left(\Delta t_{\min},\frac{\Delta t_{\max}}{\sqrt{1+\eta|E^{n}(t)-E^{n-1}(t)|^2/\Delta t_{n-1}^2}}\right),
\end{equation}
with $\Delta t_{\min}, \Delta t_{\max}, \eta$ taking $0.005, 0.1, 10^5$, respectively.
In the updating of the ascending direction $v$, the stepsize $\gamma_n$ is given by Barzilai-Borwein as \cite{Barzilai1988}
$
\gamma_n={(\Delta v^n)^\top\Delta d^n}/{(\Delta d^n)^\top\Delta d^n}
$,
where $\Delta v^n=v^n-v^{n-1}$ and $\Delta d^n=(I-(v^n)^\top v^n)\frac{(F_+^n-F_-^n)}{2l_n}-(I-(v^{n-1})^\top v^{n-1})\frac{(F_+^{n-1}-F_-^{n-1})}{2l_{n-1}}$.

To obtain the transition pathways, we start from an energy minimum $\phi_0$ and apply the SD \eqref{OSD2} to find a transition state. We choose an initial ascending direction $v^{0}$ and set the initial state of $\phi$ to be $\phi^{0}=\phi_0\pm\varepsilon v^{0}$, where $\varepsilon$ is a small positive constant. If the SD \eqref{OSD2} converges, we then find a transition state $\phi_1$ and its unstable direction $v_1$.
Next, starting from this transition state $\phi_1$, we apply the gradient flow (using \eqref{OSD2a} with $\{v_{n}\}=0$) with two initial conditions $\phi_1\pm\varepsilon v_1$ to find the two connected minima, which leads to the transition pathway between them. In most cases, these two minima are $\phi_0$ and a new local minimum, so we could confirm that $\phi_1$ is the transition state along the transition pathway connecting $\phi_0$ and the other energy minimum.

In practice, different initial searching directions are needed in order to search distinct transition states and the transition pathways connecting different local minima. Here, we choose the eigenvectors corresponding to the six smallest eigenvalues of its Hessian, as the ascending directions.

\bigskip

\section{Results}

\subsection{Bulk phases}
We first need to obtain the bulk profiles of C and G in order to formulate the C-G interface system.
We choose the coefficients in the energy \eqref{LB} such that both phases C and G are energy minima, i.e., $\xi^2=0.0389, \tau=-0.0121, \gamma=0.0681$.
These coefficients can be derived from physical parameters of diblock copolymer: volume fraction $f=0.43$ and segregation $\chi N=12$ \cite{leibler1980theory}. 
The phase profiles for both phases are discretized by $32$ grid points in the $z$-direction and $32\times 32$ Fourier modes in the $x$-$y$ plane, and they are calculated in a unit cell of the size $2\sqrt{6}\pi\times2\sqrt{6}\pi\times2\sqrt{6}\pi$, which is estimated but turns out to be very close to the optimized one for both phases.

\begin{figure}[!t]
\centering
  \includegraphics[width=0.8\columnwidth]{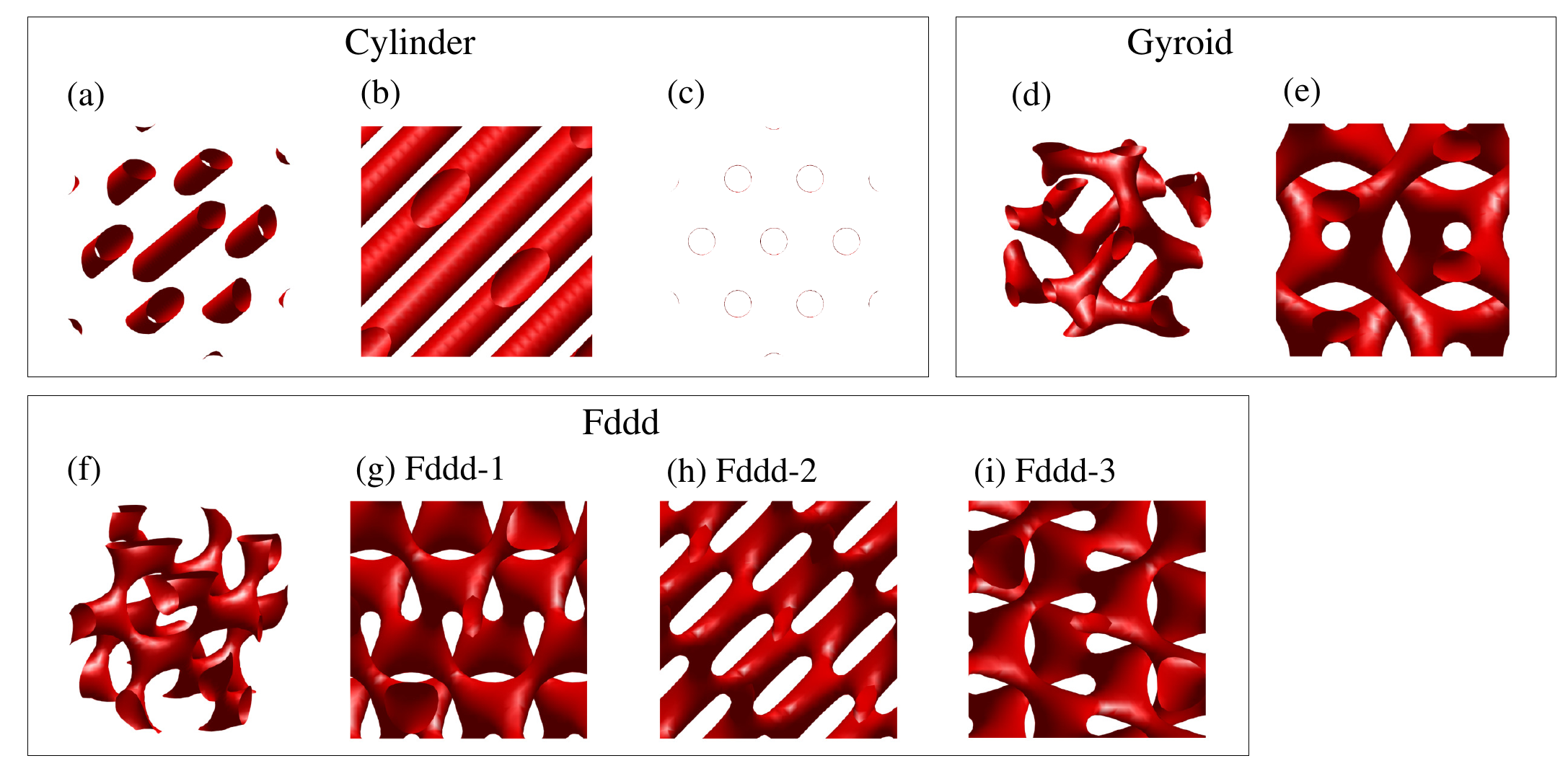}
  \caption{ C, G and Fddd phases. Specifically, (b), (e) and (g) are profiles lying in the plane ${\rm (0\bar{1}0)}$ of (a), (d) and (f), respectively; (c) is (a) lying in the plane $(111)$; (h) is (f) lying in the plane $(00\bar{1})$; (i) is obtained by rotating (g) clockwise $90^\circ$. The horizontal directions of (b), (e), (g)-(i) are $(100), (100), (100), (100), (001)$, and the vertical directions of those are $(001), (001), (001), (0\bar{1}0), (\bar{1}00)$. Here $\bar{1}$ represents its negative number $-1$.
    \label{nointer}}
\end{figure}

The configuration of C and G are presented in Figure \ref{nointer} by drawing the isosurface of their profiles.
We also show the Fddd phase, which has been reported in Refs. \citenum{Yamada2006Fddd,Kim2008Stability,Li2016Fddd} and will appear frequently in the C-G interface system.
In particular, we provide the Fddd structure viewing from three different directions, in order to be compared with the interface morphologies below, which are labelled as Fddd-1, Fddd-2, Fddd-3 to distinguish each other.
It is also notable that, under this parameter setting, G is the stable phase with $E=-1.5403\times10^{-4}$, while C and Fddd are the metastable phases with $E=-1.5190\times10^{-4}$, $-1.5186\times10^{-4}$, respectively.
The energy density of Fddd is greater than but close to C, while that of G is the lowest.

\subsection{Interface morphologies under different displacements}

We pose C and G in the orientations shown in Figure \ref{nointer}.
Specifically, C has the axis along the direction (111). The relative orientation between C and G is consistent with that in the epitaxial relation reported both theoretically\cite{Matsen1998Cylinder,Wang2011Origin} and experimentally\cite{Park2009New,Honda2006Epitaxial}.
The C-G interface system possesses three unit cells along the $z$-direction, i.e. $2L=6\sqrt{6}\pi$.
The initial condition is given by a sharp interface, with the lower one unit cell filled with the bulk profile of G, and the upper two unit cells filled with C.
By applying the gradient dynamics, the interface could be relaxed towards an energy minimum to optimize the interface.

\begin{figure}[!t]
\centering
  \includegraphics[width=0.8\columnwidth]{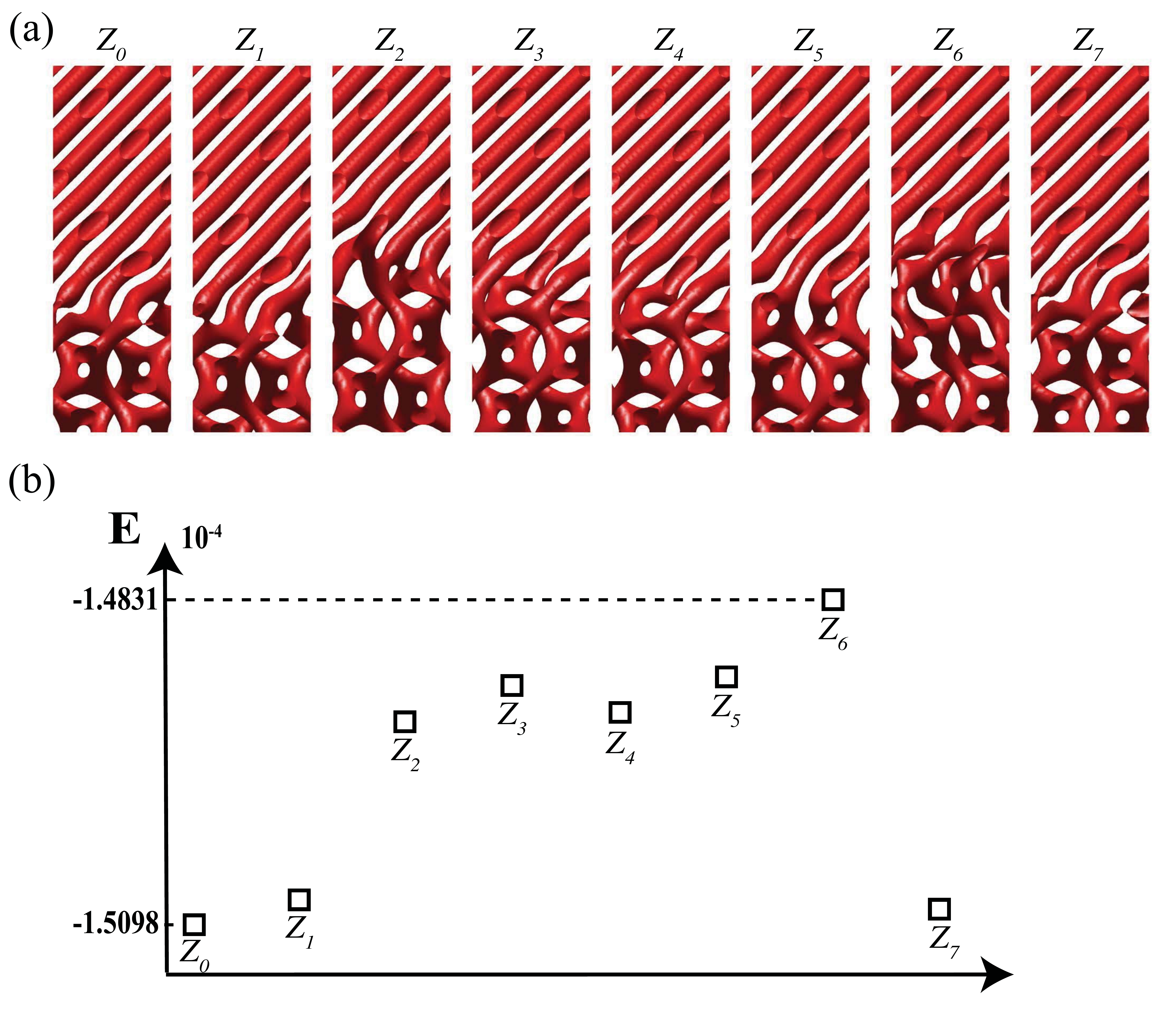}
  \caption{(a) Eight local minima $Z_i(i=0,1,...,7)$ under different displacements.
    For each $Z_i$, the corresponding initial state is set where C is fixed and G is moved down by $i$ eighths of one period. (b) The energy plot of $Z_i(i=0,1,...,7)$.
    \label{GF}}
\end{figure}

Although the orientations of C and G have been fixed, the displacement between C and G can be altered.
It turns out that the displacement makes a great difference to the interface morphology.
We investigate eight displacements along the $z$-direction and present them in Figure \ref{GF}(a).
The epitaxially matching case is labelled by $Z_0$. $Z_1$ is obtained by displacing G one eighth of the period along the $z$-direction, and $Z_2$ is the result of further displacing G one eighth of the period from $Z_1$, and so on.
While initially possessing two unit cells of C and one of G, the energies of the resulting minima, shown in Figure \ref{GF}(b), are eminently distinct due to dissimilar interface morphologies formed to connect C and G.

Let us turn to the interface morphologies.
First, the interface may prefer certain types of connections that are maintained by bulk deformation under displacements.
This is noticed in $Z_0,Z_1$ and $Z_7$, for which the connections are similar except for the interface locations and deformation degrees. As a result, their energies are quite close, and we would like to view them as the same connection mechanism.

Another mechanism is to connect two ordered phases by a third phase.
Of the eight cases displayed in Figure \ref{GF}(a), we notice that the structures of $Z_3$, $Z_4$ and $Z_6$ in the middle are not identical to either C or G.
If we only look at these local minima, one might believe that different connections could only appear under specific displacements. However, by investigating the transition pathways further, we will demonstrate that these connection mechanisms can occur under different displacements.

\subsection{Four primary transition pathways from $Z_0$}

\begin{figure}[!t]
\centering
  \includegraphics[width=0.7\columnwidth]{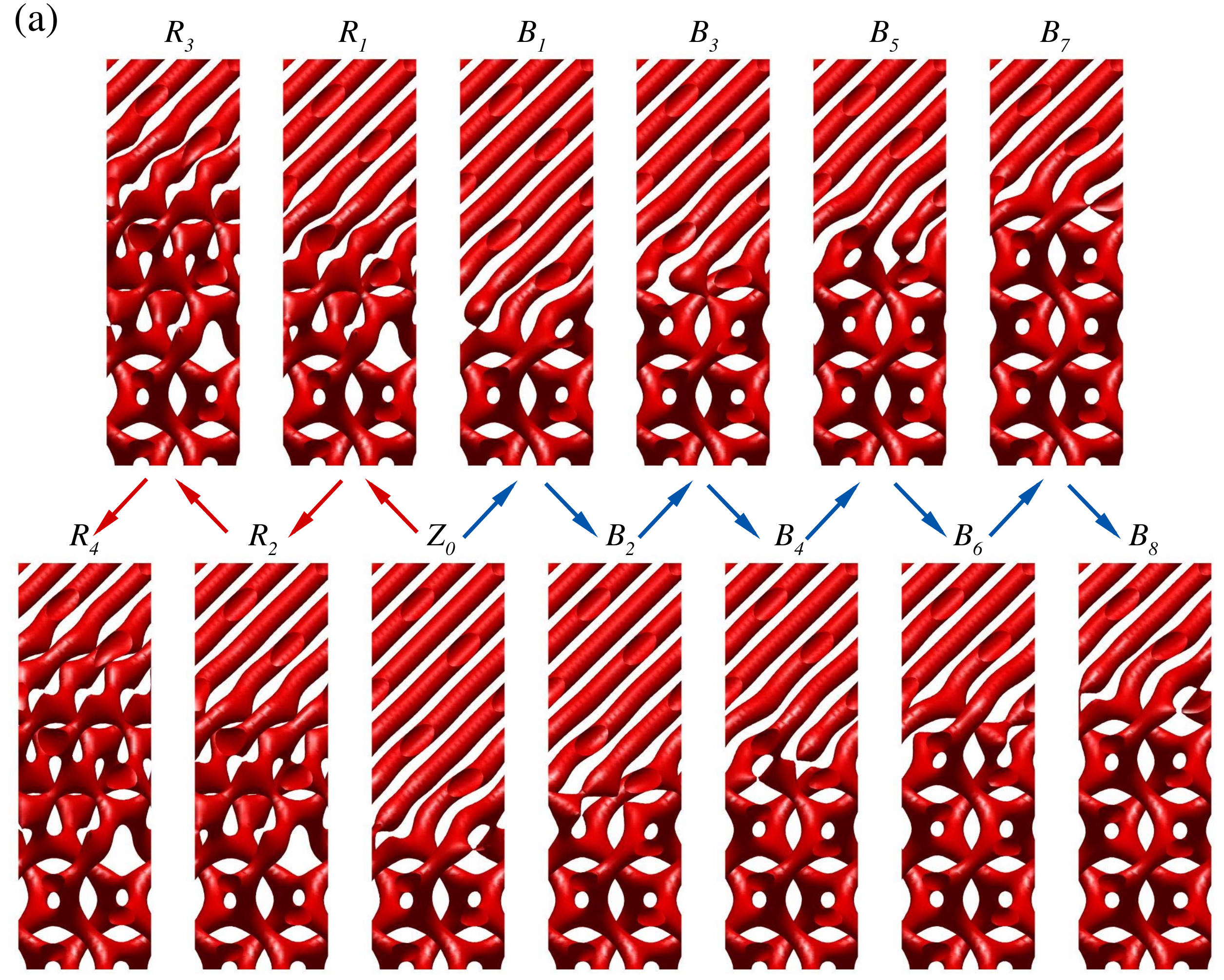}
  \includegraphics[width=0.7\columnwidth]{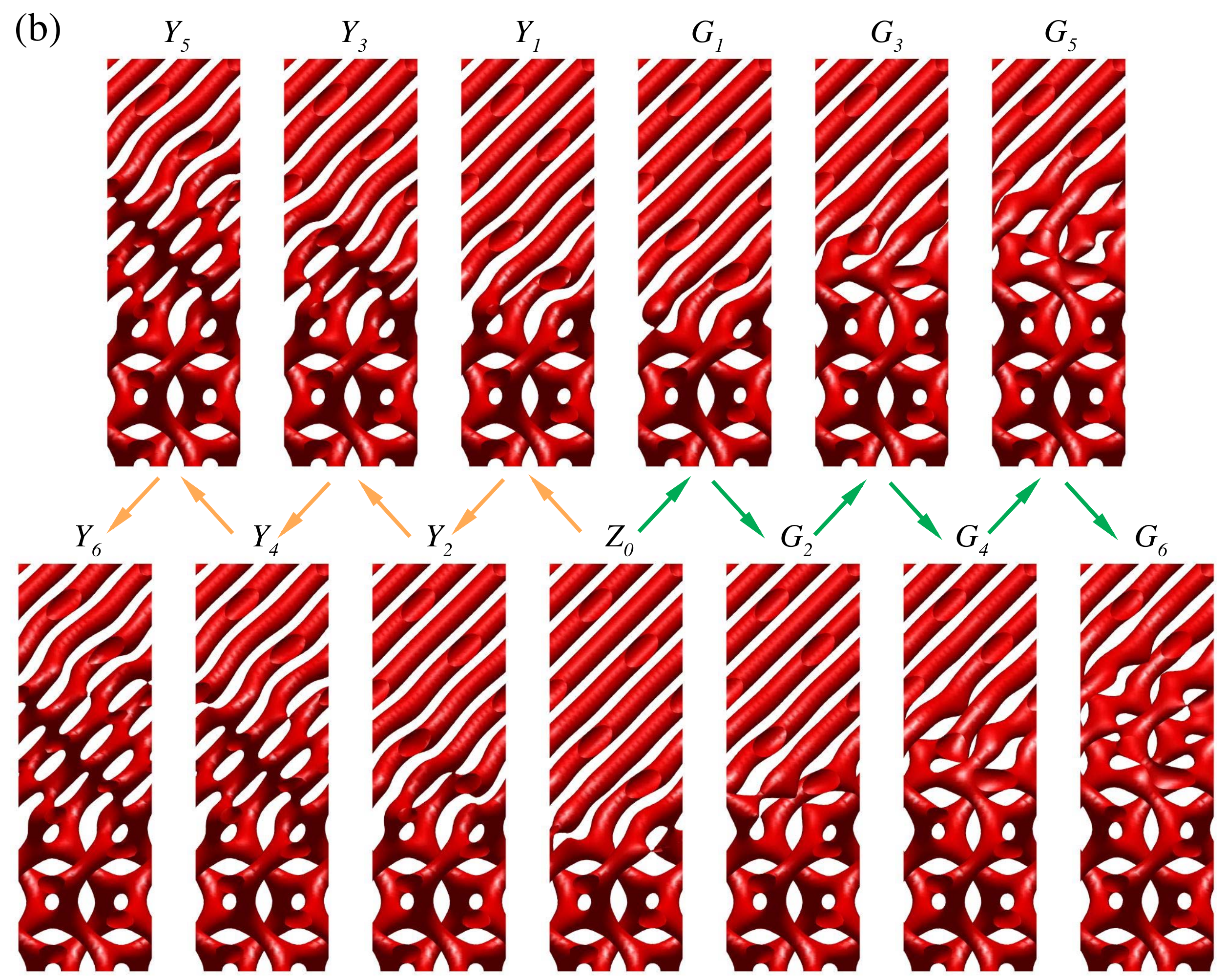}
  \caption{Four primary transition pathways starting from $Z_0$. In (a) and (b), figures in the upper row (odd subscripts) are the transition states and those in the lower row (even subscripts) are local minima. Arrows of the same color indicate a transition pathway. The figures $B_1$ and $G_1$ are identical, so are $B_2$ and $G_2$.
    \label{Z0}}
\end{figure}

\begin{figure}[!t]
\centering
  \includegraphics[width=0.7\columnwidth]{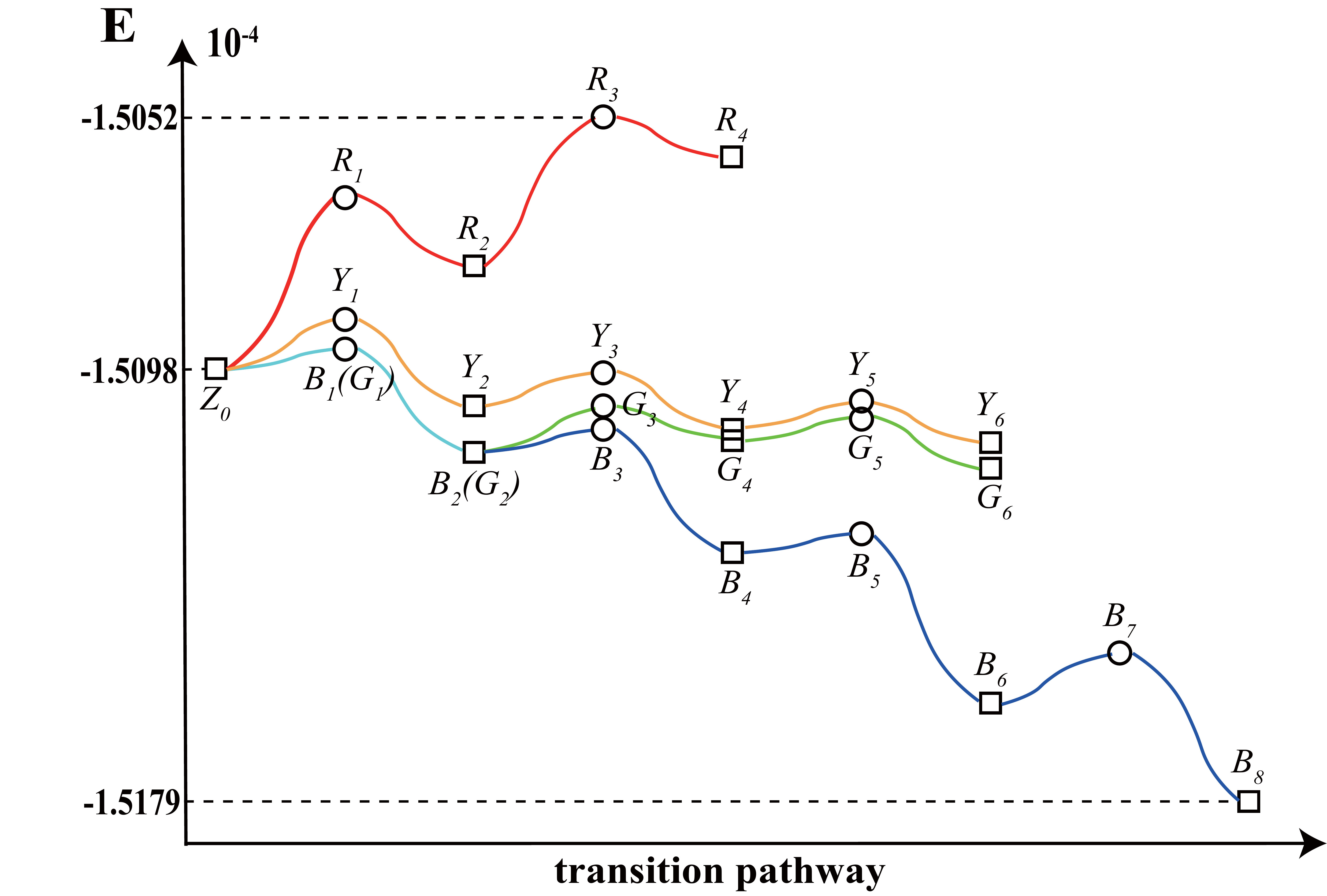}
  \caption{The free energy of four transition pathways in Figure \ref{Z0}. The circles and squares stand for transition states and local minima, respectively. The colors are the same as those in Figure \ref{Z0}. The cyan lines are the overlap of green and blue lines.
    \label{Z0e}}
\end{figure}

We first investigate the transition pathways starting from $Z_0$.
Four primary transition pathways are drawn in Figure \ref{Z0}, which we use arrows associated with different colors to represent.
We place the local minima in the lower row and the transition states in the upper row, to mimic the energy running upwards or downwards.
The corresponding free energy of each pathway is shown in Figure \ref{Z0e}.
A special notice is that part of the green and blue curves overlap, which are drawn in cyan.

In Figure \ref{Z0}, the blue arrows together compose a transition pathway showing how G is growing towards C.
The solution $B_8$ resembles $Z_0$ by moving the connection structures approximately one period towards C.
The four local minima $Z_0$, $B_2$, $B_4$, $B_6$ within one period are exactly those reported previously \cite{Xu2017Computing}.
Thus, this blue path depicts the moving of interface towards the region of C, overcoming four energy barriers in one period.
Furthermore, since the transition states are accurately identified, the energy barriers shown in Figure \ref{Z0e} are much lower than the previous findings in Ref. \citenum{Xu2017Computing}.

There are other transition pathways, where C and G are wetted by a third phase Fddd gradually.
In these cases, Fddds in red, yellow and green paths emerge in three different orientations, which resemble the three structures shown in Figure \ref{nointer} (g), (h) and (i), labelled by Fddd-1, Fddd-2, Fddd-3, respectively.
For the energy curves shown in Figure \ref{Z0e}, the green and yellow ones show a descending tendency by looking at the energy minima on the curves.
Moreover, it is noticed that $G_4$ has higher energy than $G_2$, but $G_6$ has lower energy than $G_2$ in the green curve.
The red curve exhibits an ascending tendency, which implies that this path is not favored despite its neatly grown Fddd structure.

As the blue energy curve is the lowest one in the sense that other energy curves are wholly above it, we believe that it is the most probable transition path among the four paths depicted above.
However, other paths are also possible to happen as long as their energy exhibit a decreasing tendency.
Moreover, once a state in a path is reached, it could be difficult to switch to another path.
The reason is that it might need to go back through one path to the starting point, and then move along another path, which has to overcome a higher energy barrier than keeping in the current path.

\begin{table}
\begin{center}
\caption{The main reciprocal vectors of the related bulk phases shown in Figure \ref{nointer}. $\bar{\cdot}$ represents its negative number and $(\cdot)^a$ denotes the sign of Fourier coefficients is opposite.}
\begin{tabular}{c|c}\hline\hline
Profiles  &The main reciprocal vectors\\ \hline
C & $(1\bar{2}1),(\bar{2}11), (\bar{1}\bar{1}2)$\\ \hline
G & $(\bar{1}12),(2\bar{1}1),(1\bar{2}1),(\bar{2}11),(\bar{1}\bar{1}2),(\bar{1}\bar{2}1),$\\
 &  $(112)^a,(121)^a,(211)^a,(\bar{1}21)^a,(1\bar{1}2)^a,(\bar{2}\bar{1}1)^a$\\\hline
Fddd-1 &  $(\bar{1}\bar{2}1),(\bar{1}12),(\bar{2}\bar{1}1)^a,(211)^a,(121)^a,(1\bar{1}2)^a$\\ \hline
Fddd-2 &  $(\bar{1}\bar{1}2),(121),(\bar{2}\bar{1}1)^a,(2\bar{1}\bar{1})^a,(1\bar{1}\bar{2})^a,(1\bar{2}1)^a,$\\ \hline
Fddd-3 &  $(1\bar{2}1),(211),(1\bar{1}2)^a,(11\bar{2})^a,(12\bar{1})^a,(2\bar{1}\bar{1})^a$\\
 \hline\hline
 \end{tabular}\label{table1}
 \end{center}
 \end{table}

To comprehend the wetting by the Fddd phase, we pay attention to the main reciprocal vectors of single profiles C and G, as well as Fddd in three orientations (see Table \ref{table1}).
The three main reciprocal vectors of C are part of those of G, and those of Fddd in three orientations also show resemblances.
It indicates the structural similarity between the three phases, so that the wetting by Fddd is not that surprising.
However, the spectral information is insufficient to explain the difference in energy curve.
Actually, under other displacements between C and G, we will see in the following that the red, yellow and green paths could become the one with the lowest energy.

\subsection{Transition pathways for displaced cases}
\begin{figure}[!t]
\centering
 \includegraphics[width=0.8\columnwidth]{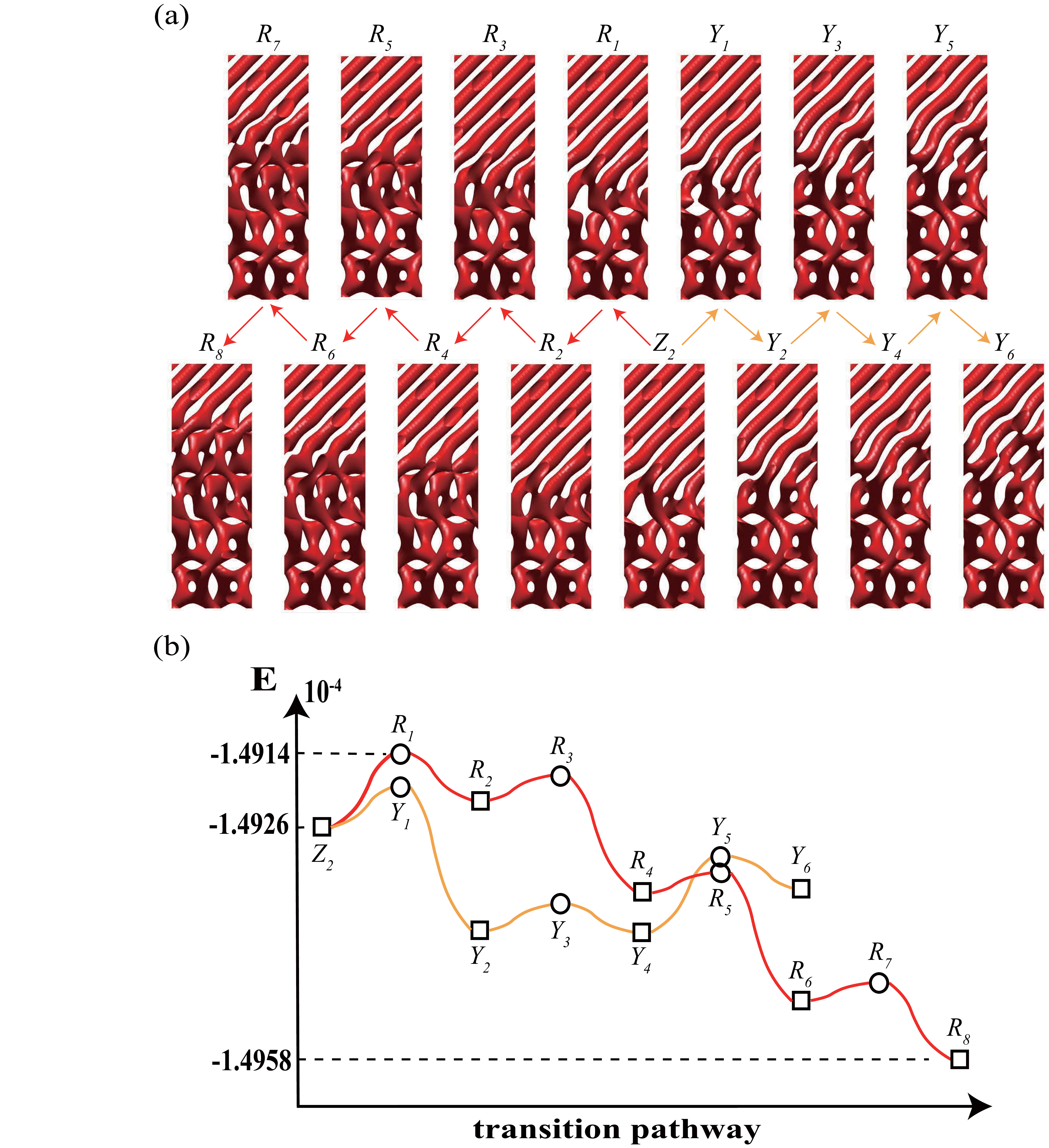}
  \caption{Two transition pathways starting from $Z_2$ are shown in (a), and the corresponding free-energy wave of each path in (a) is depicted in (b).
    \label{Z2}}
\end{figure}

Next, we will present the transition pathways starting from the other minima in Figure \ref{GF}.
In Figure \ref{Z2}(a) we find two transition pathways connecting $Z_2$, both of which are inserting Fddd.
By comparing them with Figure \ref{nointer}, we find out that along the yellow path Fddd-2 is inserted, while along the red path Fddd-1 is inserted.
We still label the solutions as $Y_1$, $Y_2$, etc., but they are not the same as those in Figure \ref{Z0} and other figures below.
Along the yellow transition pathway, $Y_2$ is at lower energy than $Z_2$, shown in Figure \ref{Z2}(b).
The Fddd structure can further grow with $Y_4$ at almost the same energy as $Y_2$, but when it grows to $Y_6$, the energy increases.
Thus, along the yellow path, the transition is likely to stop at $Y_2$ or $Y_4$.
In other words, the Fddd-2 in the middle is more like a local connecting structure than a wetting phase.
In contrast, for the red transition pathway, Fddd-1 is inserted.
The energy curve also behaves in another manner.
From $Z_2$ to $R_2$, the energy increases with a higher energy barrier than that from $Z_2$ to $Y_2$.
After that, the energy decreases notably when successive minima are reached.
This is totally different from the energy curve in Figure \ref{Z0e}, where the energy is growing when Fddd-1 is inserted into the interface.
It should be noted that the energy barrier of $Z_2\rightarrow Y_1$ is lower than that of $Z_2\rightarrow R_1$.
If we only focus on a single transition, it is easier for the transition $Z_2\rightarrow Y_2$ to happen, although the energy can reach a lower value along the red path.
Or, we could view $Y_4\rightarrow Z_2\rightarrow R_8$ as a transition pathway.
It implies that wetting by Fddd-2 needs to disappear before Fddd-1 wetting could emerge.

\begin{figure*}[!t]
\centering
 \includegraphics[width=1.0\columnwidth]{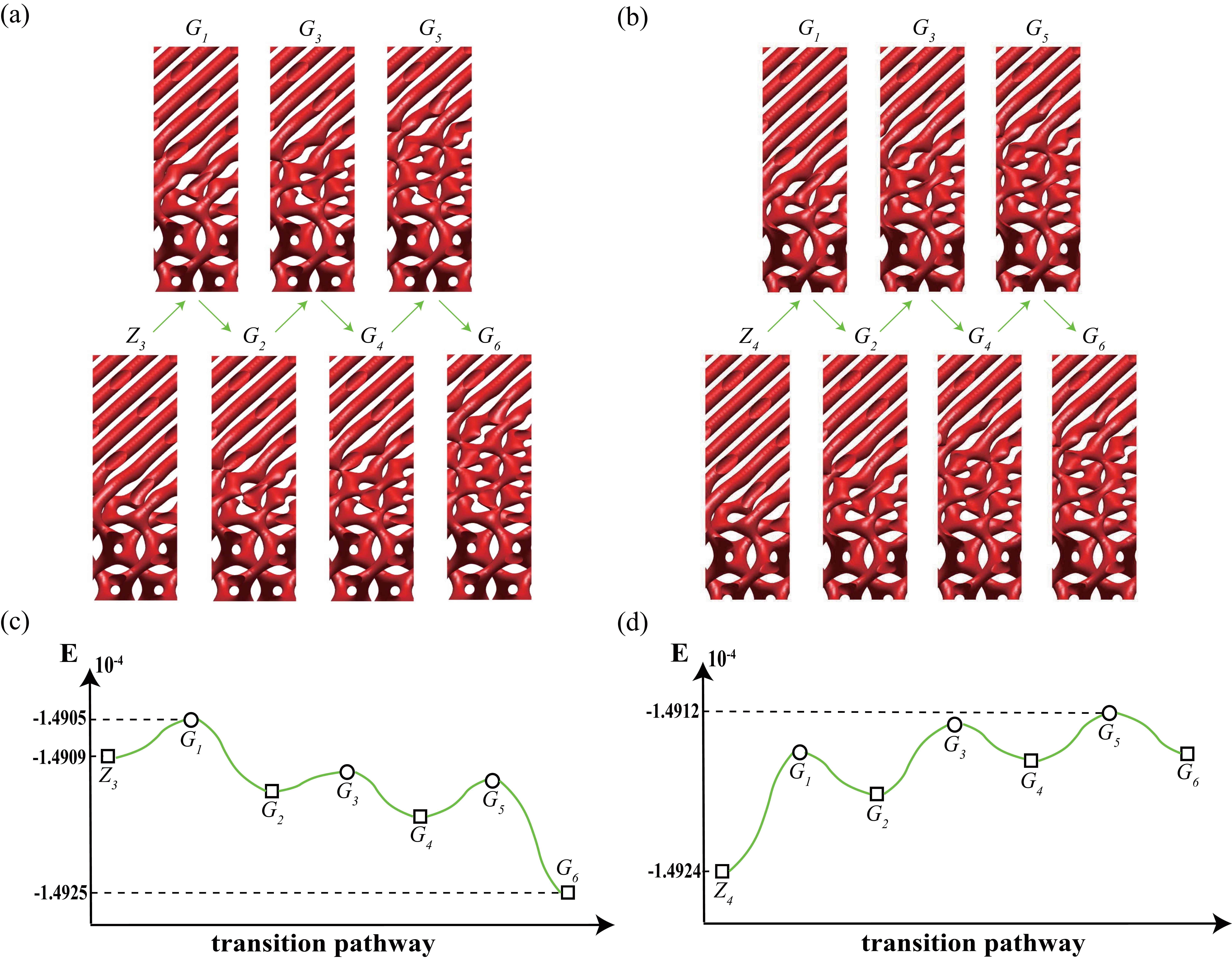}
  \caption{Transition pathways starting from (a) $Z_3$ and (b) $Z_4$ are shown, and the corresponding free-energy waves are depicted in (c) and (d), respectively. Take notice of $G_i$ starting from $Z_3$  in (a) and (c) are  different from those starting from $Z_4$ in (b) and (d).
    \label{Z34}}
\end{figure*}

We turn to the transition pathways starting from $Z_3$ and $Z_4$ (see Figure \ref{Z34}).
In both cases, we only find the paths where Fddd-3 emerges in between, so we use the color green for them.
Recall that we use the same labels to represent states in transition pathways under different displacements, so here $G_i$ in (a, c) and (b, d) are different states. 
While the connection structures for $Z_3$ and $Z_4$ are similar, their energy curves in Figure \ref{Z34}(c, d) are different.
In the $Z_3$ path, the energy is descending as Fddd-3 grows, showing that Fddd-3 is wetting.
Instead, the energy is ascending when Fddd-3 grows in the $Z_4$ path.
The results imply that $Z_4$ is the state to be stopped at and the Fddd-3 acts as a local connection.

\begin{figure*}[!t]
\centering
  \includegraphics[width=1.0\columnwidth]{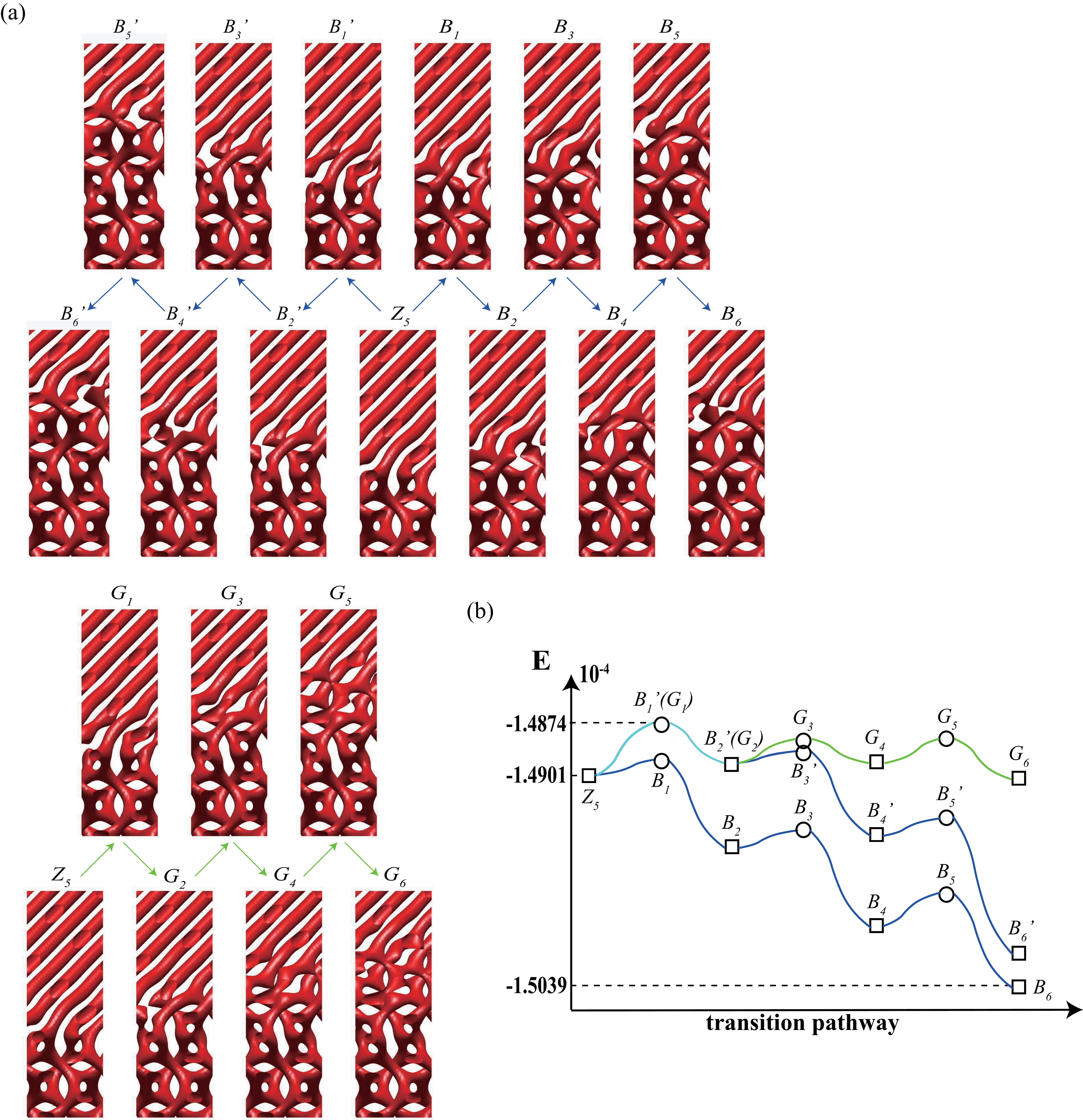}
  \caption{Three transition pathways starting from $Z_5$ are shown in (a), and the corresponding free-energy of each path is depicted in (b).
    \label{Z5}}
\end{figure*}

When starting from $Z_5$, three transition pathways are found in Figure \ref{Z5}(a).
In the two blue paths in the top ($Z_5\rightarrow B_6$ and $Z_5\rightarrow B_6'$), G is growing towards C while G is evidently deformed.
The main distinction between the two blue paths is that a twin-loop structure occurs in G in the $Z_5\rightarrow B_6'$ path.
According to the energy curve from $Z_5$ to $B_6'$ in Figure \ref{Z5}(b), the emergence of twin-loop structure draws the system to a higher free energy, while the development of G towards C makes the energy decline rapidly.
There is a green path with much higher free energy, where Fddd-3 is inserted into the interface after the emergence of twin-loop.
Similar to the viewpoint in $Z_2$, we could take $B_6'\rightarrow Z_5\rightarrow B_6$ as a transition pathway.
For the twin-loop structure, we could regard it as a sort of defect.
If such defect is formed in the interface system, the energy curve implies that it is more difficult to disappear since a much higher energy barrier $B_6'\rightarrow B_1'$ needs to be overcome.

\begin{figure*}[!t]
\centering
  \includegraphics[width=1.0\columnwidth]{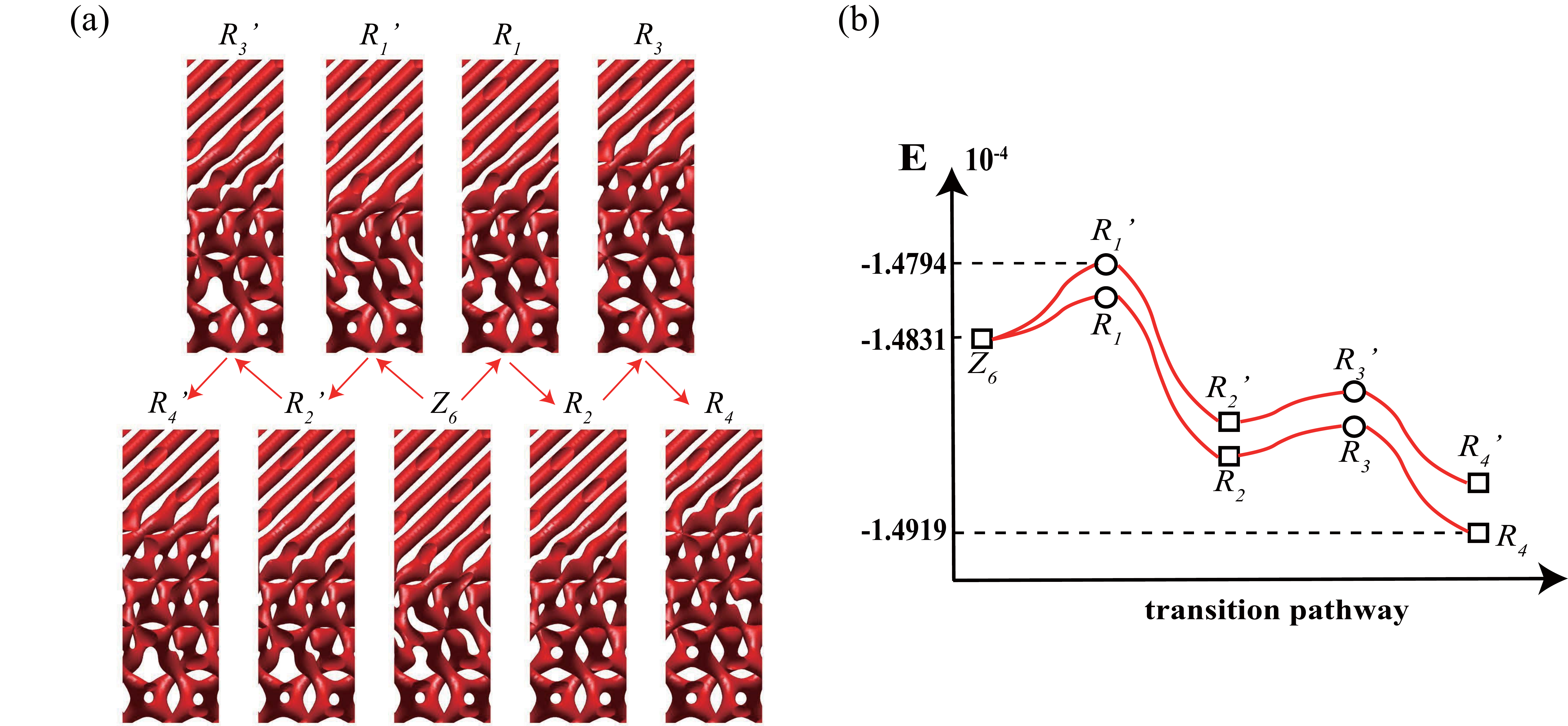}
  \caption{Two transition pathways starting from $Z_6$ are shown in (a), and the corresponding free-energy is depicted in (b).
    \label{Z6}}
\end{figure*}

At last, let us look into the transition pathways starting from $Z_6$.
The insertion of Fddd-1 is observed from two paths in Figure \ref{Z6}(a).
The difference between them is the connection between G and Fddd-1.
The path $Z_6\rightarrow R_4'$ possesses a breaking of loop in G.
The energy curves indicate that the path $Z_6\rightarrow R_4'$ has higher energy than $Z_6\rightarrow R_4$, although both paths have descending tendencies (Figure \ref{Z6}(b)).


\bigskip
\section{Discussion and conclusion}
In this work, we systematically investigate the transition pathways in the interface systems between C and G with different displacements.
We apply the SD method to efficiently compute the transition states and local minima successively to obtain transition pathways.
Application of the numerical method to the LB model reveals an interesting set of transition pathways, which describe the evolution of the C-G interfaces and reveal novel mechanisms.
We demonstrate that there exist two types of transition pathways in Cylinder-Gyroid interface: one is the direct pathway connected C and G, and the other is the indirect pathway between C and G by inserting Fddd phases with different orientations.
When C and G match well, such as $Z_0$, it is more probable for G to grow gradually towards C with notable energy decline.
However, when C or G is displaced so that the matching is broken, wetting by Fddd is favored and the orientation of Fddd show various possibilities.

By choosing the current parameter setting in the LB model, G has the lowest energy, and the energy of C is lower than Fddd but they are very close. This should be necessary for the wetting to emerge.
To connect two phases by a third phase wetting, one needs to consider the excess energy of the volume occupied by the third phase and two connections between two phases and the third phase.
If the third phase wetting is preferred, it indicates that these altogether contribute less energy than a direct connection of the two phases.
This finding is consistent with the transition pathway connecting crystals and quasicrystals, in which the lamellar quasicrystalline state serves as the third phase during the phase transition \cite{Yin2021transition}.
Thus, our results demonstrate that the flexibility of Fddd provides more possibilities on the connections between C and G phases.

Although we have identified various transition pathways in the interface system, there is no guarantee that the results are complete.
Other transition pathways might exist, showing unrealized mechanisms for the interface transition.
It may be possible to use the solution landscape approach \cite{yin2020construction, yin2021searching} for comprehensive studies of stationary points related to a high-index saddle point, which has been successfully applied to liquid crystals \cite{wang2021modelling,han2021solution,han2021a, yin2021solution} and polymers \cite{xu2021solution}. 
On the other hand, the current work only focuses on one preferred relative orientation between C and G phases. We will investigate the interface systems with other relative orientations in forthcoming works.

\begin{acknowledgement}
J. Xu was supported by the National Natural Science Foundation of China No.~12001524. L. Zhang was supported by the National Natural Science Foundation of China No.~12050002 and the National Key R\(\&\)D Program of China 2021YFF1200500.

\end{acknowledgement}

\bibliography{Inter}

\end{document}